\documentclass[12pt]{article}
\usepackage{amsmath,amssymb,graphicx}
\usepackage[compress,numbers]{natbib}
\usepackage{subfigure,placeins}
\usepackage{authblk}
\usepackage{verbatim}
\usepackage{soul} 
\usepackage{slashed}
\usepackage{floatrow}

\usepackage[font=footnotesize,labelfont=bf]{caption}

\allowdisplaybreaks
\usepackage[colorlinks=true,urlcolor=blue,anchorcolor=blue,citecolor=blue,filecolor=blue,linkcolor=blue,menucolor=blue]{hyperref}
\long\def\symbolfootnote[#1]#2{\begingroup%
\def\thefootnote{\fnsymbol{footnote}}\footnote[#1]{#2}\endgroup}

\newcommand{\newc}{\newcommand}
\newc{\gsim}{\lower.7ex\hbox{$\;\stackrel{\textstyle>}{\sim}\;$}}
\newc{\lsim}{\lower.7ex\hbox{$\;\stackrel{\textstyle<}{\sim}\;$}}
\newc{\gev}{\,{\rm GeV}}
\newc{\mev}{\,{\rm MeV}}
\newc{\ev}{\,{\rm eV}}
\newc{\kev}{\,{\rm keV}}
\newc{\tev}{\,{\rm TeV}}
\newc{\fb}{\,{\rm fb}}

\newc{\mz}{M_Z}
\newc{\mpl}{M_*}
\newc{\mw}{m_{\rm weak}}
\newc{\nr}[1]{N^c_R{}_{#1}}

\def\beq{\begin{equation}}
\def\eeq{\end{equation}}
\newcommand{\bea}{\begin{eqnarray}\begin{aligned}}
\newcommand{\eea}{\end{aligned}\end{eqnarray}}
\def\bitem{\begin{itemize}}
\def\eitem{\end{itemize}}
\newc{\ie}{{\it i.e.}}          \newc{\etal}{{\it et al.}}
\newc{\eg}{{\it e.g.}}          \newc{\etc}{{\it etc.}}
\newc{\cf}{{\it c.f.}}

 \numberwithin{equation}{section}

\newcommand\fverb{\setbox\fverbbox=\hbox\bgroup\verb}
\newcommand\fverbdo{\egroup\medskip\noindent%
            \fbox{\unhbox\fverbbox}\ }
\newcommand\fverbit{\egroup\item[\fbox{\unhbox\fverbbox}]}
\newbox\fverbbox

\newcommand{\be}{\begin{equation}}
\newcommand{\ee}{\end{equation}}

\newcommand{\alb}[1]{{\color{red} #1}}

\begin{document}

\author[1]{
Mihailo Backovi\'c\thanks{mihailo.backovic@uclouvain.be,}}
\author[2]{Alberto Mariotti\thanks{alberto.mariotti@vub.ac.be}}
\author[3,4]{Diego Redigolo\thanks{dredigol@lpthe.jussieu.fr}}
\small
\affil[1]{\small{Center for Cosmology, Particle Physics and Phenomenology - CP3,
Universite Catholique de Louvain, Louvain-la-neuve, Belgium}
}
\affil[2]{ \small{Theoretische Natuurkunde and IIHE/ELEM, Vrije Universiteit Brussel,
and International Solvay Institutes, Pleinlaan 2, B-1050 Brussels, Belgium}}
\affil[3]{Sorbonne Universit\'es, UPMC Univ Paris 06, UMR 7589, LPTHE, F-75005, Paris, France
}
\affil[4]{CNRS, UMR 7589, LPTHE, F-75005, Paris, France
}

\title{Di-photon excess illuminates Dark Matter}

\maketitle

\vspace*{-10cm}
\noindent 
CP3-15-46
\vspace*{9cm}

\begin{abstract}
We propose a simplified model of dark matter with a scalar mediator 
to accommodate the di-photon excess recently observed by the ATLAS and CMS collaborations.
Decays of the resonance into dark matter can easily account for a relatively large width of the scalar resonance, while the magnitude of the total width combined with the constraint on dark matter relic density lead to sharp predictions on the parameters of the Dark Sector.
Under the assumption of a rather large width, the model predicts a signal consistent with $\sim 300 \text{ GeV}$ dark matter particle in channels with large missing energy. This prediction is not yet severely bounded by LHC Run I searches and will be accessible at the LHC Run II in the jet plus missing energy channel with more luminosity.
Our analysis also considers astro-physical constraints, pointing out that future direct detection experiments will be sensitive to this scenario.

\end{abstract}

\newpage


\section{Introduction}\label{Intro}
Both ATLAS and CMS collaborations recently announced a search for reasonances in the di-photon channel, featuring an excess of events around $m_{\gamma \gamma}\approx$ 750 GeV \cite{atlas, CMS:2015dxe}. The result sparked an enormous interest within the theoretical physic community, with no less than 35 preprints appearing by the end of the week from the LHC and ATLAS announcement \cite{Harigaya:2015ezk,Mambrini:2015wyu,Franceschini:2015kwy,DiChiara:2015vdm,Angelescu:2015uiz,Pilaftsis:2015ycr,Ellis:2015oso,Bellazzini:2015nxw,Gupta:2015zzs,Molinaro:2015cwg,Higaki:2015jag,McDermott:2015sck,Petersson:2015mkr,Dutta:2015wqh,Cao:2015pto,Matsuzaki:2015che,Kobakhidze:2015ldh,Cox:2015ckc,Ahmed:2015uqt,Agrawal:2015dbf,Martinez:2015kmn,No:2015bsn,Demidov:2015zqn,Chao:2015ttq,Fichet:2015vvy,Curtin:2015jcv,Bian:2015kjt,Chakrabortty:2015hff,Csaki:2015vek,Falkowski:2015swt,Aloni:2015mxa,Bai:2015nbs}.

The excess suggests a resonance structure ($X$) with:
\begin{align}
&\sigma(pp\to X)\times  \text{BR}(X\to\gamma\gamma) \gtrsim 2 \text{ fb}\label{feature1}\ , \\
&m_X \approx 750 \text{ GeV}\label{feature2},
\end{align}
where $X$ is the new resonance singly produced in $pp$ collisions.
An interesting feature of the di-photon excess seems to be the moderately large total width of the resonance. While the CMS  collaboration does not have enough events to provide information on the width of the resonance, the best fit of the ATLAS collaboration favors a rather large width of around $6\%$ of the  resonance mass. For the purpose of this paper we are going to assume 
\begin{equation}
\frac{\Gamma_{\mathrm{tot}}(X)}{m_X}\approx \text{3-9} \% \ . \label{feature3}
\end{equation}

 The appearance of the di-photon excess impels some effort to relate the origin of the fluctuations to some well-motivated new physics scenarios.  
The Landau-Yang theorem \cite{Landau:1948kw,Yang:1950rg} forbids direct decays of an on-shell spin-1 particle into di-photons, implying that the new physics models which can explain the excess contain likely either CP-even $(S)$, CP-odd $(a)$ scalars, or a spin two massive object $(G^{'\mu\nu})$. In both scenarios the coupling to photons is controlled by operators of dimension five or higher,  and hence generically suppressed with respect to tree level couplings by at least one power of a new mass scale and some loop factors. 

While a spin 2 object can certainly be motivated in the context of extra dimensions \cite{Giudice:1998ck}, in this paper we focus on the simpler possibility of a scalar resonance which we take to be a singlet of the Standard Model. Extra singlets are indeed ubiquitous in many scenarios of Beyond the Standard Model (BSM) physics, motivated either by naturalness considerations \cite{Chacko:2005pe,Ellwanger:2009dp} or by dark matter physics \cite{Chu:2012qy,Kim:2008pp,LopezHonorez:2012kv,Farina:2013mla}.

Taking the large width constraint \eqref{feature3} at face value, it is rather difficult to satisfy it by assuming a dominant branching ratio into photons, essentially because of the dimensional suppression of the scalar couplings to photons.
Moreover, increasing the width via decays into SM pairs is likely to be challenged by exclusion limits on new resonances from the LHC Run~I at $\sqrt{s}=7$ and 8 TeV. 

The lack of observation of either new charged states with mass $< m_{S,a} / 2\sim 375 \gev$, or excesses in other SM channels which suggest a 750 GeV resonance, invites us to consider the possibility that the large width of the singlet resonance is a result of decays into a Dark Sector. The singlet could then play the role of  a ``scalar mediator'' between the Standard Model and the Dark Sector where, say,  a fermionic candidate for dark matter (DM) resides. We will show how requiring that the fermionic DM candidate fully accommodates the observed relic abundance of $\Omega_{DM}h^2\approx0.12$ \cite{Ade:2013zuv} determines completely the parameters of the Dark Sector, namely the DM mass and its coupling strength to the singlet. 

Even if we will focus our discussion on the large width scenario \eqref{feature3}, the same framework can certainly accomodate the di-photon excess in the narrow width scenario.  This is the case in regions of the parameter space where the DM mass is $ \gtrsim m_S/2$,  as well as in regions featuring very small couplings between the singlet and the DM, which naturally lead to a small decay width for $S$. However, we find such scenarios less motivated, since the introduction of a DM candidate in this case does not serve a purpose of explaining any feature of the di-photon signal.

The possibility of a new scalar resonance as the mediator between the Dark Sector and the Standard Model (SM) is very much in the spirit of simple models of singlet DM \cite{Chu:2012qy, Kim:2008pp,LopezHonorez:2012kv,Farina:2013mla}. In order to simplify the discussion we are going to focus on a minimally model dependent case of a CP-even scalar $S$ coupled to a Dirac-like fermonic dark matter. 

We present a simple scenario where the di-photon excess is a manifest of the singlet scalar DM portal with effective couplings to SM gauge bosons (while the Higgs portal coupling is negligible). The observed features of the di-photon excess combined with cosmological constraints on DM  lead to a sharp prediction of a dark matter candidate with mass $m_{DM} \sim 300 \gev \lesssim m_S/2 ,$ and a coupling to the scalar of $O(1)$. 

This prediction is consistent with the existing experimental bounds from correlated LHC-8TeV searches like $\gamma Z$ \cite{Aad:2013izg}, di-jets \cite{Khachatryan:2015sja} and $j+$MET \cite{Khachatryan:2014rra} and from dark matter direct and indirect  detection experiments. Moreover the sizeable coupling of the singlet to dark matter suggests that the LHC-13TeV searches for large missing energy associated with a jet, $\gamma$, $h$ or $Z$ should observe a signal consistent with dark matter of $m_{DM}\approx300 \gev$ and a mediator with mass $m_{S} \approx 750 \gev$ in the near future. Furthermore, such signal will be within the reach of the next generation direct detection  \cite{Cushman:2013zza} experiments.

From the point of view of the ultra-violet (UV) completions of our simplified model, a challenge will be to suppress the mixing of the singlet with the SM Higgs and at the same time have sizeable couplings to photons and gluons and to DM pairs.  In the appendix \ref{UVcompl} we discuss a concrete UV complete scenario in the context of supersymmetry (SUSY) where the pseudo-modulus (sgoldstino) generically associated to spontaneous SUSY-breaking can be responsible for the DM portal \cite{Shih:2009he,KerenZur:2009cv,Amariti:2009tu} and have the right structure of the couplings. Motivated by SUSY scenarios (where a the pseudo-modulus is in general a complex scalar) we comment on how the phenomenology will change in the presence of a CP-odd singlet $a$. A more comprehensive analysis is left for future works.

The remainder of the paper is organised as follows. In section \ref{sec3} we present our simplified model for Dirac dark matter based on a CP-even scalar portal. We  discuss its decay channels and how the requirements on the di-photon resonance and dark matter can be fulfilled. In section \ref{secExp} we discuss the experimental constraints
from the LHC-8TeV and from DM detection experiments and present a final summary of the allowed parameter space.
In section \ref{sec4} we discuss the prospects for future detection of our DM candidate both at the LHC-13TeV and the next generation direct detection experiments. Appendix \ref{UVcompl} contains a discussion about the possible UV completions of the simplified model presented in section \ref{sec3} and a brief analysis on the phenomenology of a CP-odd singlet motivated by SUSY UV completions.



\section{Di-photon excess  in a dark matter simplified model}\label{sec3}
We consider an effective lagrangian for a new spin zero and CP-even particle $S$ $(J^p=0^+)$ which couples at tree level to a massive Dirac fermion $\psi$. 
Both $S$ and the fermion are singlet under the Standard Model and a 
global flavor symmetry under which $\psi$ is charged guarantees a stable fermionic DM candidate
\begin{align}
\mathcal{L}_{NP}^+&=\frac{1}{2}(\partial S)^2+\frac{m_S^2}{2} S^2+\bar \psi\slashed{\partial}\psi +
(g_{DM} S+M_{\psi})\bar\psi \psi\notag\\
&+\frac{g_{GG}}{\Lambda}S G^{\mu\nu} G_{\mu\nu}+\frac{g_{WW}}{\Lambda}S W^{\mu\nu} W_{\mu\nu}+\frac{g_{BB}}{\Lambda}S B^{\mu\nu} B_{\mu\nu}\ .\label{lagrangian}
\end{align}
We fix the UV scale to $\Lambda=10^4$ GeV conservatively sticking to the regime of validity of the effective field theory. 
The dimensionless couplings are taken to be order $O(1)$, while
the missing operators allowed by the symmetries in the effective lagrangian are assumed to be suppressed by small couplings.
Here we focus on the basic phenomenological properties of the simplified model of dark matter in (\ref{lagrangian}), 
taking a bottom-up approach\footnote{For our phenomenological studies, we employ FeynRules \cite{Alloul:2013bka}, MadGraph5 \cite{Alwall:2011uj}, 
MadDM \cite{Backovic:2013dpa,Backovic:2015cra} and micrOMEGAs \cite{Belanger:2013oya}.}, while we postpone  the justification of our working assumptions in considering (\ref{lagrangian}) for the Appendix \ref{UVcompl}.

\begin{figure}[h!]
\includegraphics[width=0.5\textwidth]{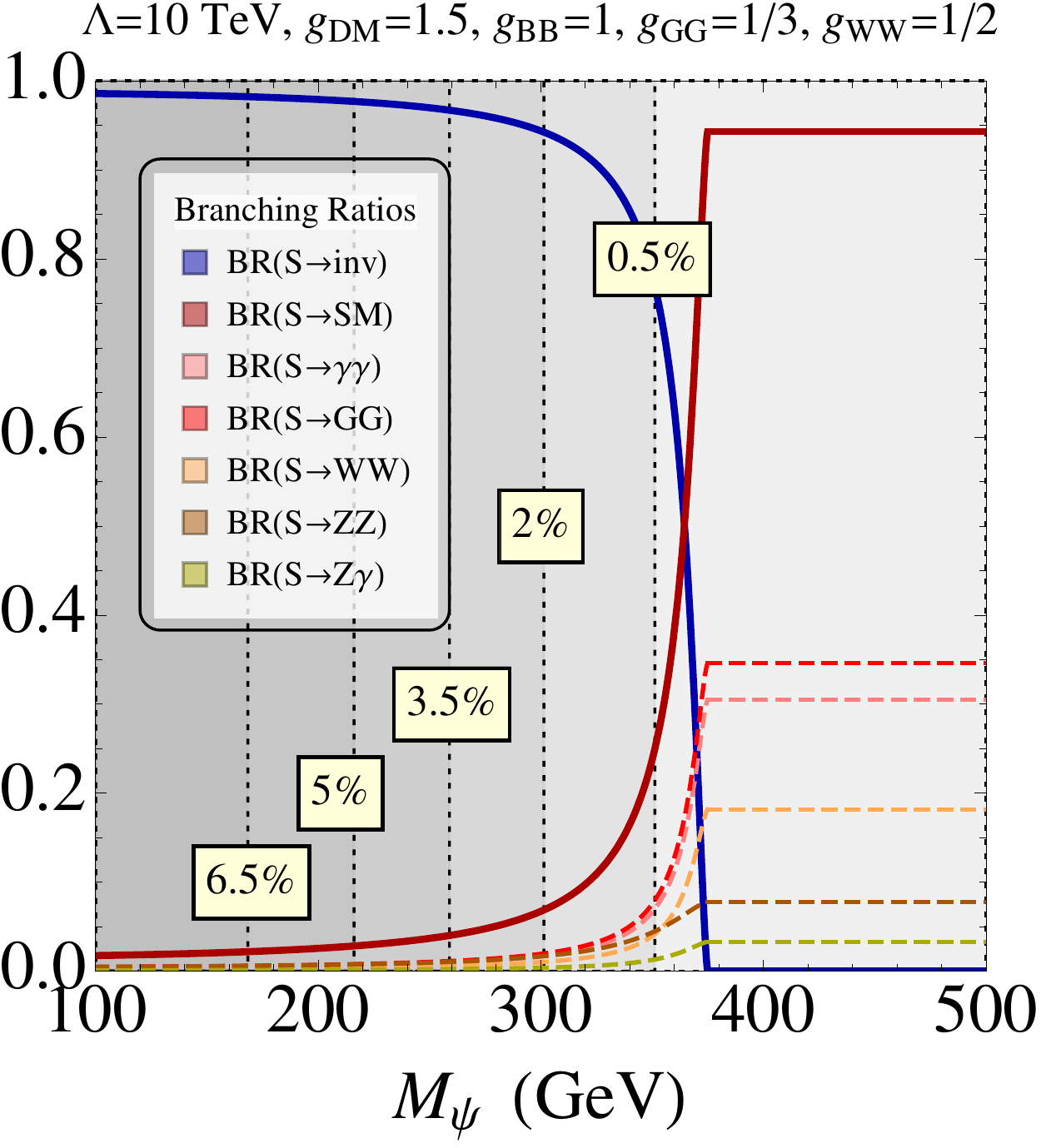}\\
\caption{Contours of the different branching ratios for the singlet $S$ as a function of $M_\psi$. The other parameters of the model are fixed to the benchmark values specified in the plot title. The solid lines indicate the $\text{BR}$ to invisible (blue) and SM particles (red). The dashed lines are the different SM channels. The gray shaded contours indicate  $\Gamma_{tot}/m_S$ where we fixed $m_S=750\text{ GeV}$.}\label{BRplot}
\end{figure}

\paragraph{The total width:} Considering the couplings in (\ref{lagrangian}), the singlet scalar can decay into SM gauge bosons,  or invisibly with the leading order decay rate
\be
\Gamma(S \to \bar \psi \psi) = \frac{g_{DM}^2 m_{S}}{8 \pi }\left(1-\frac{4 M_{\psi}^2}{m_{S}^2} \right)^{3/2}\label{invisibleCPeven}\ .
\ee
In figure \ref{BRplot} we display the branching ratios of $S$ decays into the various channels, for some representative values of the couplings,
as a function of the DM mass. As soon as the tree level decay into dark matter is kinematically open, it dominates over the decays into SM particles which are induced by dimension five operators. Among the SM decay channels, the gluon decay mode is enhanced by the color factor.

Figure \ref{BRplot} illustrates an important feature of the model. The gray-shaded contours indicate the ratio of the width of $S$ over its mass, i.e. $\frac{\Gamma_{tot}}{m_S}$.
As we pointed out in the introduction, the ATLAS analysis
hints towards a configuration of the spectrum and the couplings for which $\frac{\Gamma_{tot}}{m_S} \sim \text{3-9} \%$.
Figure \ref{BRplot} clearly shows the difficulties in obtaining a percent-level width by considering dominant decay modes into SM particles, which contribute $\lesssim 0.5\%$ to $\frac{\Gamma_{tot}}{m_S}$. This feature is generic of models where the decay modes into SM particles are generated through higher dimensional operators only, if we conservatively stick to the regime of validity of the effective field theory. In such scenarios a tree level decay mode certainly helps to enhance the width of the resonance. 

Figure \ref{BRplot} also shows that a large width can generically be obtained via the invisible decay of the singlet into DM pairs \eqref{invisibleCPeven}. Indeed the requirement on a large width alone \eqref{feature3} imposes a lower bound on the Yukawa-like coupling of the singlet to DM: $g_{DM}>1$. In what follows we will show how this leads to very interesting implications for both DM direct detection as well as collider phenomenology.

\begin{figure}[h!]
\includegraphics[width=0.6\textwidth]{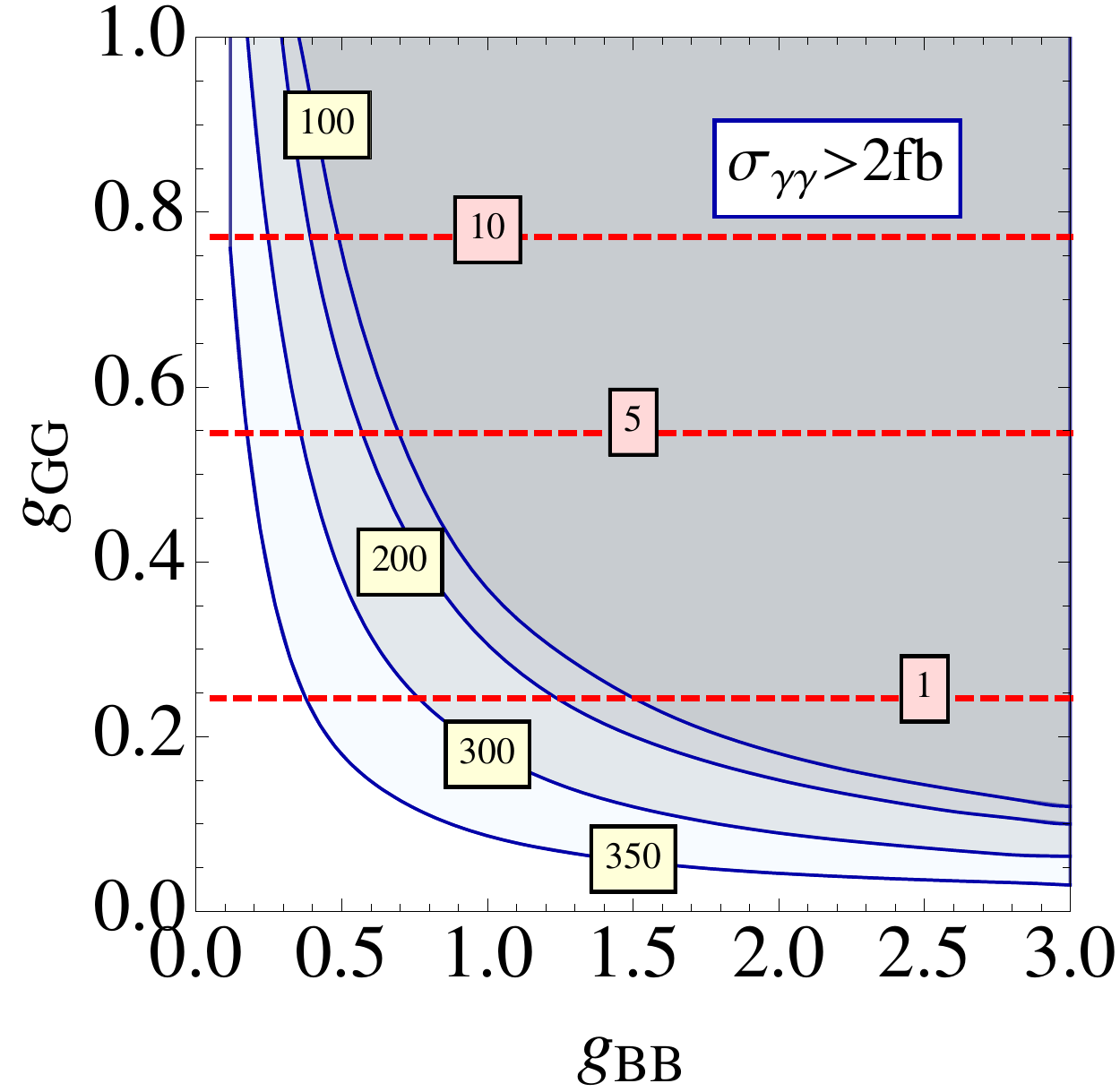}
\caption{Parameter space (from blue to gray) in the plane of couplings $(g_{BB},g_{GG})$ leading to $\sigma(pp\to S\to\gamma\gamma)>2\text{ fb}$ at $\sqrt{s}=13$ TeV, for different choices of the DM mass $M_\psi=(350,300,200,100)\text{ GeV}$, indicated on the boundaries of the regions. For the purpose of illustration, we fixed  $g_{DM} = 2.7$ and $\Lambda = 10 \tev$. The red dashed contours correspond to the production cross section at 13 TeV for a scalar singlet $\sigma(pp\to S)$ in pb. 
}
\label{xsecplot} 
\end{figure}

\paragraph{The di-photon signal strength:} The new singlet scalar $S$ can be produced via gluon fusion at the LHC through the dimension five operator controlled by $g_{GG}$. Once the UV cut-off $\Lambda$ is fixed the production cross section is determined by $g_{GG}$. Figure \ref{xsecplot} shows that by dialing $g_{GG}$ one can easily achieve $\sigma(pp\to S)\approx\mathcal{O}(\text{pb})$. 

The branching ratios into SM channels are fixed by the parameters $g_{WW}, g_{ZZ}$ and $g_{GG} $ in the dimension five operators in \eqref{lagrangian} and result in decay modes which contribute to di-jet, di-boson, di-photon and $Z\gamma$ final states. In order to reduce the number of parameters of the model we will fix $g_{WW} \approx0$ in the remaining of this paper.  Note that suppressing $g_{WW}$ is going to artificially enhance the $\gamma \gamma$ channel with respect to $Z Z$ and $W W$. However, modifying this assumption will not affect our main conclusion. Assuming $g_{WW} \approx0$  the $\gamma \gamma$ signal cross section scales like $\sigma(X\to\gamma \gamma) \sim g_{GG}^2 g_{BB}^2$ at fixed dark matter mass and $g_{DM}$.

In figure \ref{xsecplot} we also give an idea of the expected signal strength, before cuts, for a scalar singlet $S$ decaying into photons at LHC-13TeV as a function of the DM mass which we take to be below $\frac{m_S}{2}$ so that $S$ has always a sizeable invisible width \eqref{invisibleCPeven}.

The regions where $\sigma(S\to\gamma\gamma)>2\text{ fb}$ depends on the value of the DM mass $M_\psi$ once the coupling strength $g_{DM}$ is fixed to be $\mathcal{O}(1)$. At fixed dark matter mass, the boundary of the viable signal strength region reproduces the expected parametric dependence on $g_{GG}^2 g_{BB}^2$.  The maximum signal strength is achieved when the DM mass approaches the kinematical limit $M_\psi\sim\frac{m_S}{2}$ and the invisible width controlled by \eqref{invisibleCPeven} is reduced. 

From figure \ref{BRplot} and figure \ref{xsecplot} we see an interesting tension between enhancing the $\gamma \gamma$ signal strength and the total width at the same time. A large cross section in $\gamma\gamma$ would prefer a DM mass close to the kinematic threshold in order to suppress $\sigma(S\to \mathrm{invisible})$. On the other hand, a large width of order $\frac{\Gamma_{tot}}{m_S} \sim \text{few} \%$ prefers a  DM mass of $O(100)$ GeV. In section 4 we will see how these two constraints can be put together the LHC bounds selecting a specific region of the $g_{GG}$-$g_{BB}$ plane where also a viable DM candidate can be accommodated. 


\paragraph{Dark matter relic abundance:} The model we discuss also aims to account for the observed relic abundance of DM in the Universe, i.e. $\Omega h^2 \simeq 0.12$.
The annihilation cross section of DM into SM particles is driven by higher dimensional operators, typically resulting in  annihilation rates which are too low to obtain the correct $\Omega h^2$ for generic values of the dimensionless couplings and dark matter mass.
The correct value for $\Omega h^2$ can be obtained if the annihilation is kinematically enhanced, i.e. if the mass of the DM is ``close'' \footnote{While the term ``close to the resonance'' is often used in the context of resonant dark matter annihilation,  it is seldom pointed out that the relevant measure for correct relic density is in fact  $|m_S - 2M_\Psi| / \Gamma_{tot} \lesssim O(1)$.} to the singlet resonance. Note that this is exactly the same region where the signal strength in $\gamma\gamma$ is maximized as shown in figure \ref{xsecplot}. In the same region the kinematical suppression reduces the invisible width of the scalar \eqref{invisibleCPeven} and hence the total width (see figure \ref{BRplot}). We then expect the Dark Matter mass and coupling  $g_{DM}$ to be fully determined by the intersection of the relic density and resonance width constraints. 

\begin{figure}[t]
\includegraphics[width=0.6\textwidth]{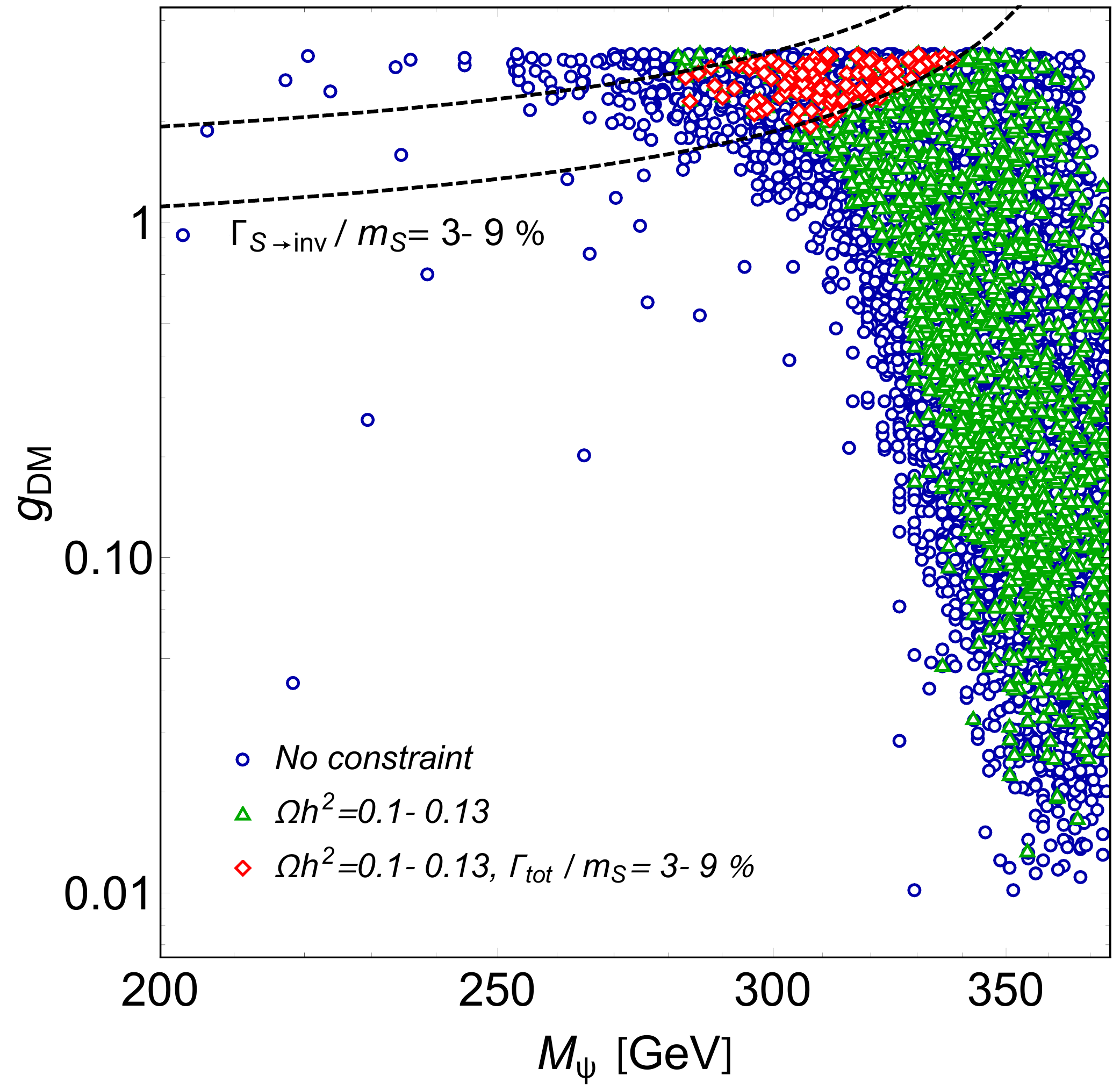}
\caption{Markov-chain scan over the four dimensional model parameter space, projected onto the $M_{\psi}, g_{DM}$ plane. The scan assumes a Gaussian likelihood function centered around $\Omega h^2 = 0.12, $ where the range of allowed parameters is bounded by $g_{BB} = [10^{-2}, 2], \,g_{GG} = [10^{-2}, 1], \,g_{DB} = [10^{-2},3]$ and $m_\psi = [200, 375]$ GeV. The blue circles represent the total of 10000 points scanned over by the Markov-chain, with no additional constraints. The green triangles represent a subset of the sampled points which give relic density in the range of $0.1 < \Omega h^2 < 0.13$. The red diamonds assume an additional requirement of $\tfrac{\Gamma_{tot}}{m_S} = (3 -9)\% \gev$. The dashed lines represent the range in which the total width in the range of $(3-9)\%$ of $m_S$ can be explained by dominant decays into dark matter. 
\label{fig:markov}}
\end{figure}

Indeed, requiring a large width in combination with relic density alone provides strong constraints on the parameters of the Dark Sector. 
In order to clarify this aspect further, we performed a Markov-chain exploration of the model parameter space in $\{ g_{GG}, g_{BB}, g_{DM}, M_{\psi} \}$. Figure \ref{fig:markov} shows the results,  where we projected the four dimensional parameter space onto the $(g_{DM}, M_{\psi})$ plane.
The correct relic density and a large total width can be obtained only in the region of  dark matter mass of the order of $M_{\psi} \sim 300$ GeV, regardless of the values of $g_{BB}$ and $g_{GG}$. In the following we will show that some of the model points which satisfy both the relic density and the large width requirement (red diamonds in figure \ref{fig:markov}) are also able to accommodate the observed di-photon signal strength. 

The exact desired values of the Dark Sector will depend on the value of the total width and somewhat on the spin of the DM particle and the chiral nature of the di-photon resonance. However, the overall conclusion that the requirement of a total width combined with relic density will essentially fix the parameters of the Dark Sector appears to be robust and weakly dependent on the remaining model parameters.
 \\

\paragraph{Benchmark points:} \begin{figure}[h!]
\includegraphics[width=0.46\textwidth]{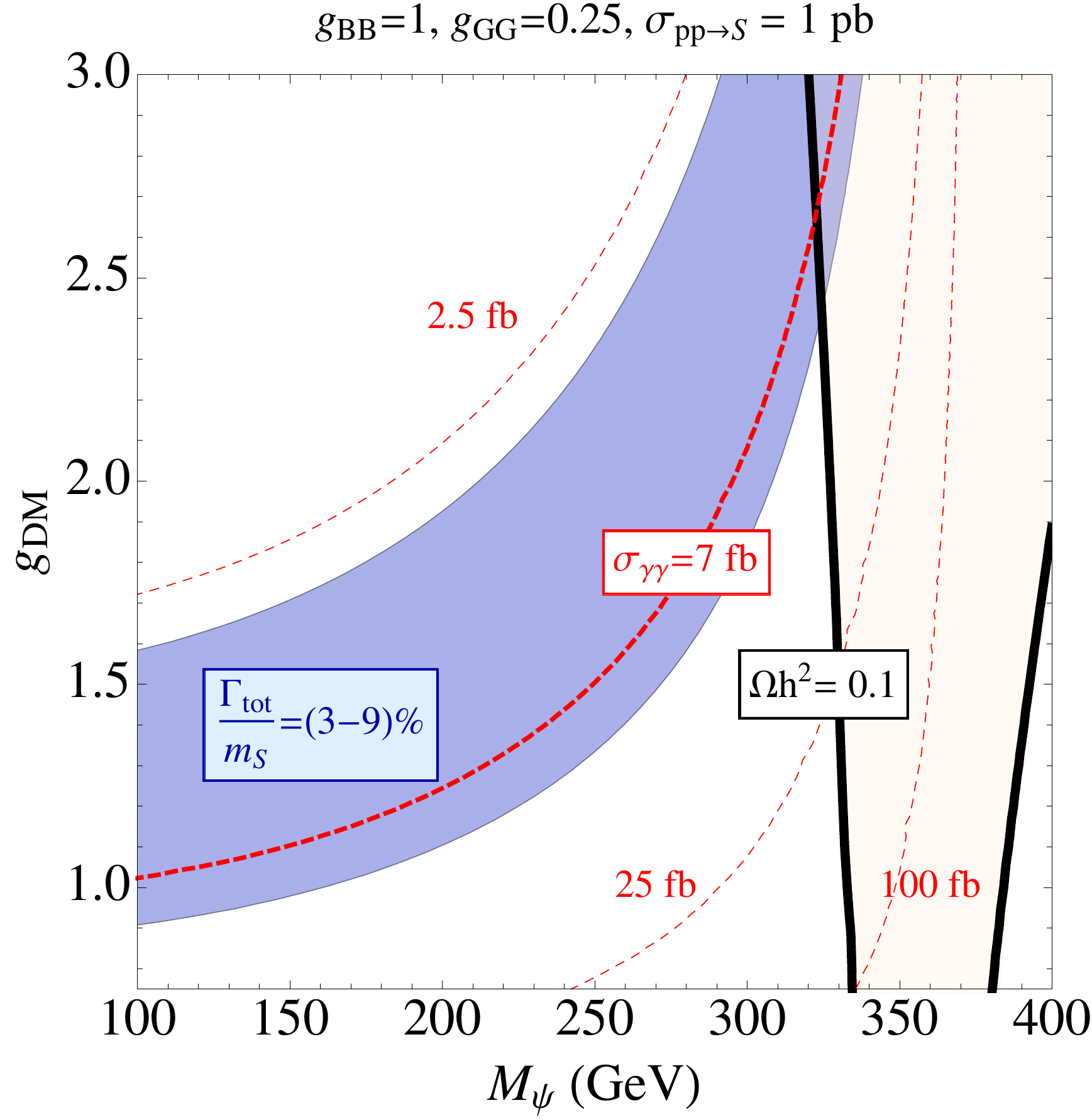}
\hspace{0.5cm}
\includegraphics[width=0.46\textwidth]{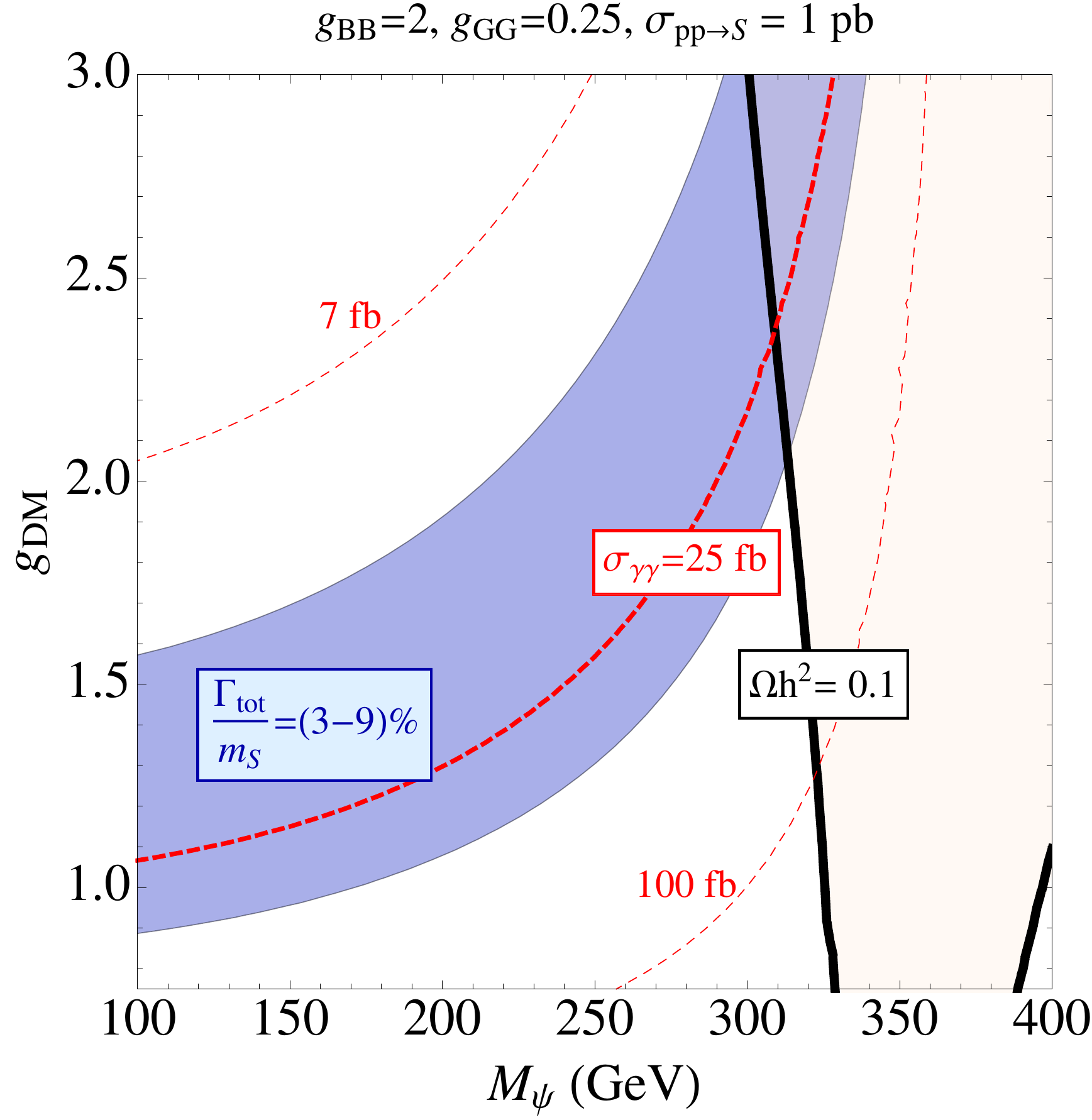}
\includegraphics[width=0.46\textwidth]{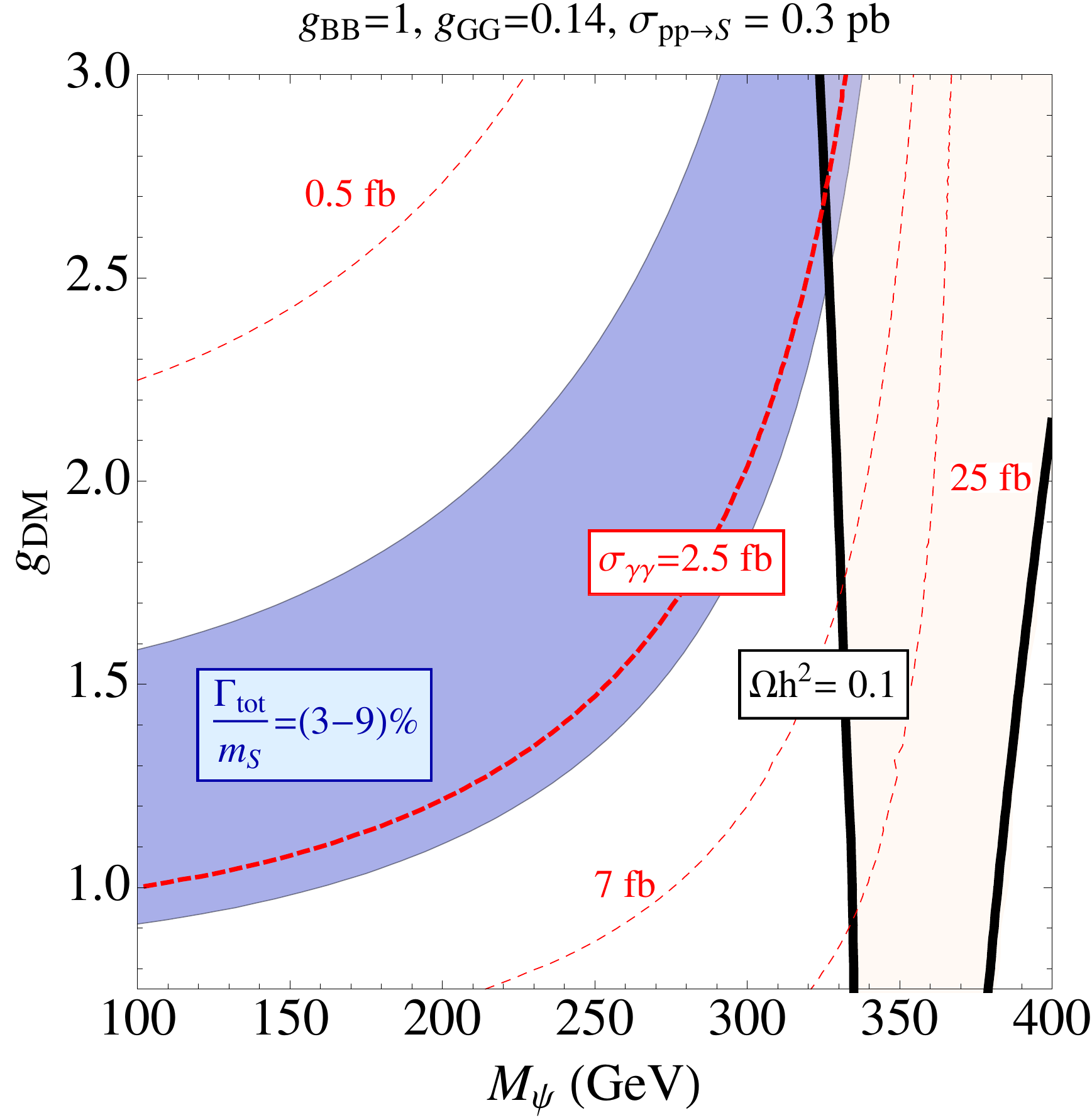}
\hspace{0.5cm}
\includegraphics[width=0.46\textwidth]{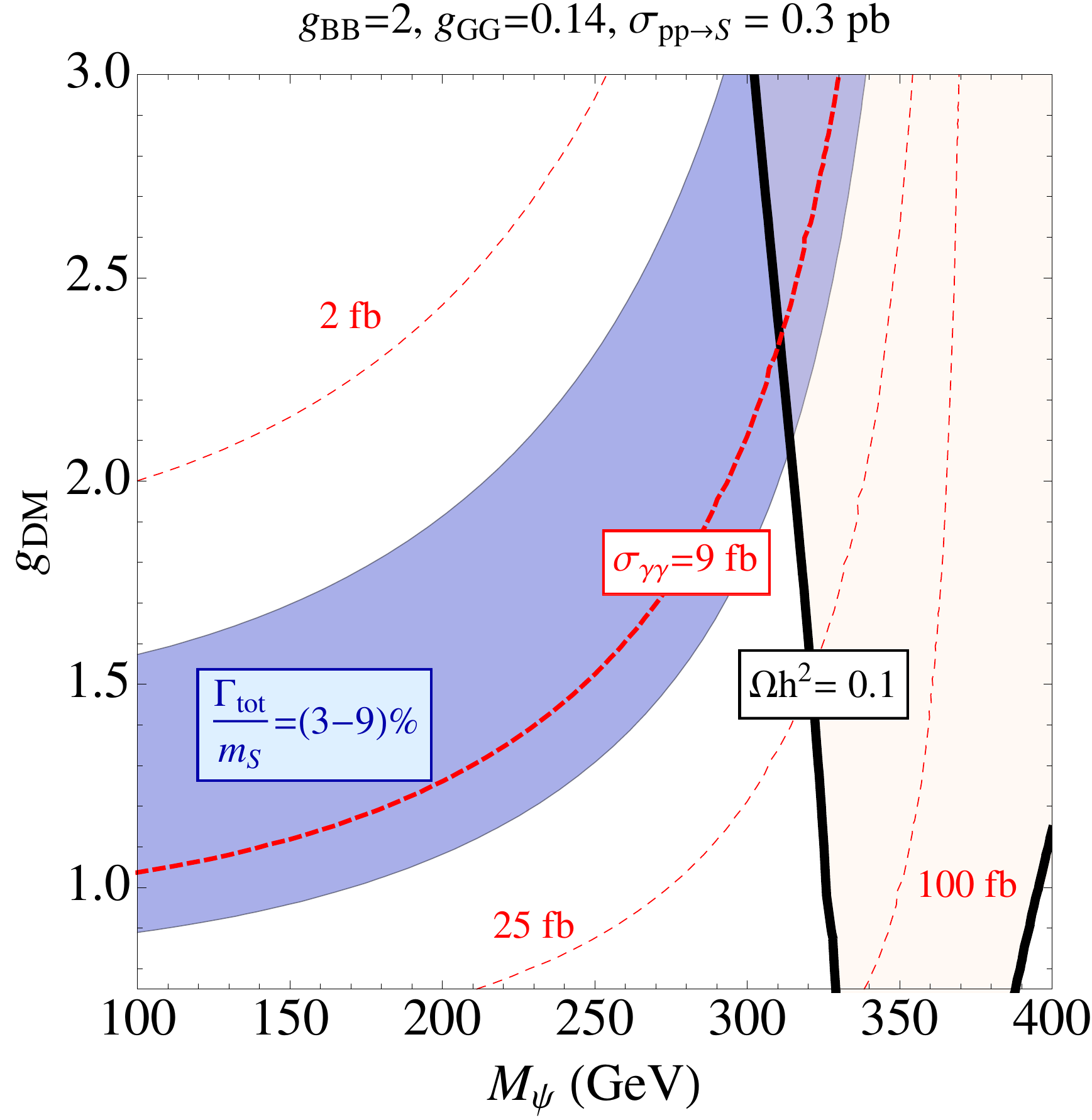}
\caption{Interplay between the di-photon signal parameters and DM relic density. The four plots from the top left to the bottom right corresponds to our four benchmark choices in \eqref{benchmarks}. The blue band shows regions where $\frac{\Gamma_{tot}}{m_S} \approx (3-9)\%$ (in blue) while the region where $\Omega h^2<0.1$ is shown in pink.  At the boundaries of the pink region (black, solid lines) the  Dirac fermion accounts for all the DM relic abundance. We overlaid the contours of the di-photon production cross section at LHC 13 as dashed red curves. The thicker curve is the one corresponding to our selected benchmark with $\Gamma_{tot}\approx30\text{ GeV}$. Notice that the regions where the width bands intersect the regions allowed by relic density constrains essentially fix the dark matter parameters.}
\label{fig:sexy}
\end{figure}

For concreteness we selected four benchmark points which provide a yield in $\gamma \gamma$  of $O(1-10)$ fb, roughly required to explain the observed di-photon excess (\ref{feature1}):

\bea
&
\text{p}_1 : \qquad g_{GG}=0.25 \qquad g_{BB}=1 \, ,\\
&
\text{p}_2 : \qquad g_{GG}=0.25 \qquad g_{BB}=2 \, , \\
&
\text{p}_3 : \qquad g_{GG}=0.14 \qquad g_{BB}=1 \, ,\\
&
\text{p}_4 : \qquad g_{GG}=0.14 \qquad g_{BB}=2 \,  ,
\label{benchmarks}
 \eea
where we are keeping fixed the cut-off scale at $\Lambda = 10\, \tev$.
We intentionally choose $\mathcal{O}(1)$ values for  $g_{BB}$, which opens up the parameter space leading to a sizeable $\gamma \gamma$ cross section.

Figure~\ref{fig:sexy} illustrates our main results for the four benchmark points defined in \eqref{benchmarks}.
For illustrative purpose, we select a large range for the mediator width \eqref{feature3} with the ATLAS preferred value of $\Gamma_{tot}/m_S=6\%$ as central value.

The panels in the figure show where the band with $\Omega_{DM}h^2<0.12$ overlays the band where the width of the scalar mediator $S$ is in 
the preferred range.
The relic abundance band is weakly dependent on the coupling as expected for an annihilation cross section dominated by a an $s$-channel resonance and fixes the Dark Matter mass to be $M_{\psi} \sim \frac{m_S}{2}$. For the large values of the dark matter mass necessary to obtain the correct relic abundance, a width in the selected range can be achieved only with a large dark matter coupling $g_{DM}$. This result is in agreement with the expectation that the dark matter relic abundance together with the requirement on the width of $S$ essentially fixes both $g_{DM}$ and $M_{\psi}$.

By inspecting figure \ref{fig:sexy}, we can select representative values of the dark matter mass and dark matter coupling 
$g_{DM}$, where we chose the total width $\Gamma_{tot}\sim 30$ GeV for illustrative purpose.
Table \ref{tab:fit} summarizes the signal yield in the di-photon channel and the DM relic abundance for the four selected benchmark points, where all the parameters of the model are now fixed by requiring a large mediator width, a sizable $\gamma \gamma$ cross section, and the correct dark matter abundance.

Note that the $\gamma \gamma$ production cross section for our benchmarks ranges between $2 - 25$ fb, providing enough room to fit the ATLAS and CMS excess while taking into account event selection efficiency and the acceptance.

 \begin{center}
\begin{table}
\footnotesize
\begin{tabular}{ccccccc}
   benchmark  & $(g_{GG} , g_{BB})$ & $g_{DM}$ & $M_{\psi}$ (GeV)  &$\Gamma_{tot}$(GeV) &$\sigma_{\gamma \gamma}$(fb) at 13TeV &$\Omega h^2$ \\ 
      \hline
      \hline
$p_1$&(0.25,1)& 2.7 & 322& 30 & 6.2  & 0.10 \\
$p_2$&(0.25,2)& 2.2& 307& 29& 25   & 0.12 \\
 $p_3$&(0.14,1)& 2.7 & 323& 29& 2.1  & 0.12 \\
$p_4$&(0.14,2)& 2.3& 308& 31& 7.8   & 0.12 \\
\hline
\end{tabular}
\caption{Summary of the di-photon signal yield at LHC-13TeV and the preferred dark matter parameters for the four selected benchmark points.
}
\label{tab:fit}
\end{table}
\end{center}

\section{Experimental constraints}\label{secExp}
The scenario we consider is bounded by several existing collider searches at $\sqrt{s}=8$ TeV and by astro-particle searches that we discuss in more detail in the following sections. For a previous study of LHC Run I constraints on DM models with mediators see e.g. \cite{Jaeckel:2012yz}.
\paragraph{LHC-8TeV constraints:} 
The model we propose populates different final state topologies, given the rich decay pattern of the scalar mediator (see figure \ref{BRplot}).
For the benchmark points that we selected in order to accommodate the di-photon excess as well as to obtain the correct relic abundance 
the largest contribution to collider signals is in channels with missing energy. However, the channels in which the mediator decays into SM particles, even if suppressed
by a small branching fraction, can also lead to stringent bounds.
%


The most relevant collider bounds from LHC searches at $\sqrt{s}=8 \tev$ are:
\begin{itemize}
\item Recent CMS \cite{Khachatryan:2015qba} search for a di-photon resonance in the mass range $150$ to $850$ GeV. 
For a scalar resonance with mass $m_S \sim 750$ GeV, their results impose an upper bound of $\sigma_{\gamma \gamma} \lesssim 2$ fb.
\item The ATLAS \cite{Aad:2014fha}  measurement of the $Z\gamma$ final state places a bound of $\sigma_{\gamma Z} \lesssim 3.5$ fb for scalar resonances of $m_S\sim 750 \gev$.
\item Mono-jet searches provide the most stringent bounds on signals with large missing energy. CMS \cite{Khachatryan:2014rra} as well as ATLAS 
\cite{Aad:2015zva} put a bound of $\sigma(\mathrm{MET+X}) \lesssim 6 \,  \mathrm{fb}$  for signals with $\mathrm{MET} > 500 \gev$.
\item Recent CMS di-jet searches for resonances at $\sqrt{s} = 8 \tev$ \cite{Khachatryan:2015sja} provide weak limits for production cross section of $\sigma(jj)  \lesssim 1 $ pb for scalar resonances which couple dominantly to $gg$ of mass around the TeV scale. 
We will adopt this limit for a scalar resonance of $m_S \sim 750 \gev$ as a conservative estimate. 
\item The ATLAS search \cite{Aad:2015kna}
 provides a bound on the $ZZ$ cross section of the order $\sigma_{ZZ}~<~12$~fb for a scalar resonance of $m_S \sim 750$ GeV.
In our scenario the $ZZ$ cross section is suppressed with respect to the $\gamma \gamma$ cross section
by a factor $( \frac{s_W}{c_W} )^4 \sim 0.1$,  since we fixed $g_{WW}\approx0$. Hence the bound on $\sigma_{ZZ}$ is less relevant than the ones on $\sigma_{\gamma\gamma}$ and $\sigma_{Z\gamma}$.

\end{itemize}

\begin{figure}[h!]
\includegraphics[width=0.6\textwidth]{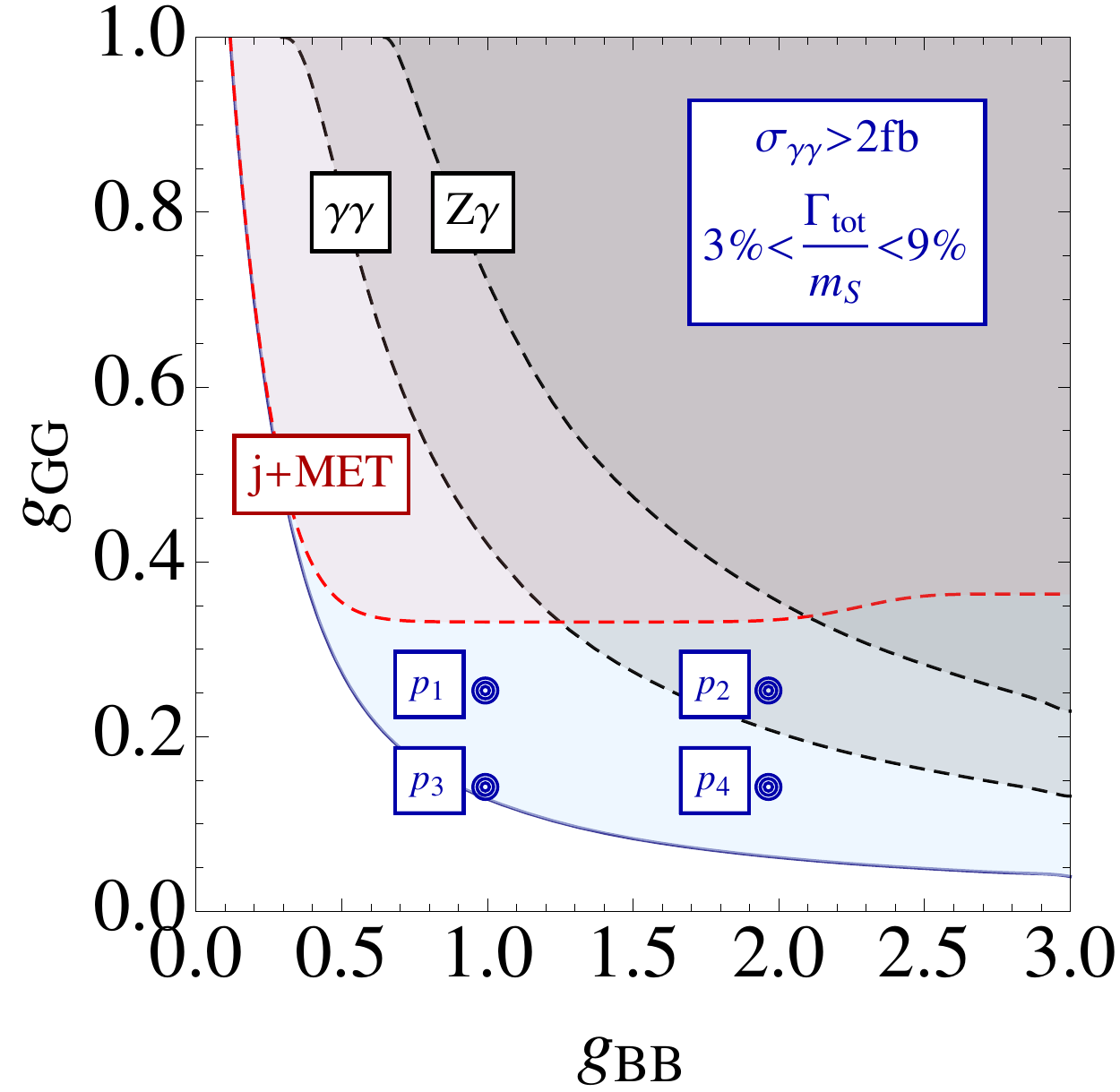}\\
\caption{
Summary of the parameter space allowed by collider constraints,  in the $g_{GG}$ and $g_{BB}$ plane. We marginalise over $M_{\psi}=[0,600] \text{ GeV}$ and $g_{DM}=[0,3]$. Regions above dashed lines are ruled out by individual collider searches specified on the plot. The solid blue line corresponds to the region of parameter space which can provide  $\sigma(pp\to S\to\gamma\gamma)>2\text{ fb}$ with dark matter mass of 300 GeV. The blue points labeled as $p_{1-4}$ represent the benchmark model points we use as illustrations in the paper. 
}
\label{xsecplot2} 
\end{figure}

\noindent
%
The above list summarizes the relevant LHC-8TeV constraints.
The question of how much of the parameter space which can explain the di-photon signal strength is still allowed by the 8 TeV collider searches remains.
For this purpose, we performed a scan over $\{ g_{GG}, g_{BB}, g_{DM}, M_{\psi} \}$, accepting only points 
featuring a width in the range $3\% \leq \frac{\Gamma_{tot}}{m_S}\leq 9\%$.  Figure \ref{xsecplot2} shows the results,
where we projected the four dimensional scan 
onto the $(g_{GG}, g_{BB})$ plane. 
Signal yield of $\sigma_{\gamma\gamma} \gtrsim 2 $ fb can be obtained only in the region above and to the right of the solid blue line (as expected since the $\gamma\gamma$ cross section scales like $g_{GG}^2 g_{BB}^2$ for a fixed total width).

In the same plot we display the bounds from LHC-8TeV searches as dashed lines. 
The exclusion lines in figure \ref{xsecplot2} are not accounting for further constraints on $M_\psi$ and $g_{DM}$ coming from having the right DM relic abundance. The latter forces the DM mass to be around $300$ GeV slightly suppressing the invisible decay and hence making the LHC bounds more severe.  

Our results show that a portion of parameter space with  $g_{GG} \lesssim 0.3$, assuming $g_{BB} \sim 1$,  is still allowed by the LHC-8TeV data. Note
that this region of model parameters is also able to accommodate the di-photon excess signal strength.
The four benchmark points \eqref{benchmarks} thoroughly investigated in the previous section
are labeled with $p_{1-4}$. 
The figure suggests that 
the second benchmark point is already excluded with the LHC-8TeV $\gamma \gamma$ constraint,  
but we anyway use it as an example
which can lead to a large signal yield.\\

\paragraph{Dark matter detection constraints:}
Beside collider bounds, we expect that our dark matter model can also be constrained by direct and indirect detection experiments. 

Direct detection experiments can constrain the model since the lagrangian (\ref{lagrangian})
induces the following effective operator between the dirac dark matter and the gluons 
\be
\mathcal{L}_{eff} \supset \frac{ g_{DM} g_{GG}}{\Lambda m_S^2} \bar \psi \psi G_{\mu \nu} G^{\mu \nu} .
\ee
Notice that the strength of this operator is correlated with the requirement on the large total width as shown in figure \ref{fig:sexy}. The resulting spin independent cross section for DM scattering off nucleons is then given by (see e.g. \cite{Chu:2012qy})
\be
\label{DDsigma}
\sigma_{SI}^{(p,n)} = 
\frac{1}{\pi} \frac{(m_{\chi} m_N)^2}{(m_{\chi}+m_N)^2} \left( m_{p,n}\frac{g_{DM}g_{GG}}{\Lambda m_S^2} f^{(p,n)}_G \right)^2,
\ee
where $f^{(p,n)}_G=\frac{8 \pi}{9 \alpha_s}  \left( 1-\sum_{q=u,d,s} f^{(p,n)}_q  \right)$ is the gluon form factor and $\alpha_s$ is the evaluated at the scale of the singlet mass. In our estimate of the direct detection constraints we are neglecting subleading operators which will be generated by the running from the UV scale to the typical scale of direct detection experiment $(\approx \text{GeV})$. This operators should be added in a more precise treatment of direct detection bounds.\footnote{We thank Paolo Panci for interesting discussions on this point.} 
The LUX experiment \cite{Akerib:2013tjd} provides a limit on the contact interaction between scalar mediators and gluons of $\sigma_{SI} \lesssim 
4 \times 10^{-45}$ cm$^2$ for a dark matter of mass around $300$ GeV.

Concerning indirect detection, the annihilation is velocity suppressed in the case of a real scalar mediator. We  hence do not
expect strong bounds on our model from measurements of galactic gamma ray fluxes. We regardless  estimate  the cross section for annihilation of galactic DM into photons in our benchmark points for completeness.
Recent measurements of galactic gamma rays from the FERMI collaboration \cite{Ackermann:2015lka} put a bound of $\langle \sigma v \rangle_{\gamma \gamma} \lesssim 10^{-28} \frac{\text{cm}^3}{\text{s}}$ for a DM mass of $O(300)$ GeV that we adopt for our scenario.\footnote{The bound depends on the halo profile and varies in the range $(10^{-27}-10^{-28}) \frac{\text{cm}^3}{\text{s}}$.}\\

\paragraph{Benchmark points:}
Table \ref{tab:constraints} shows a summary of all the experimental constraints on our scenario for the four benchmark model points. 
Benchmark point $2$, with $(g_{GG},g_{BB})=(0.25,2)$, gives the largest yield in the di-photon signal (see Table \ref{tab:fit}) and it is already severely constrained by the $\gamma \gamma$ final state, as already shown in figure \ref{xsecplot2}. Interestingly, requiring the correct DM relic abundance for that choice of $g_{GG}$ and $g_{BB}$ point enhances the $Z\gamma$ branching ratio making the benchmark 2 also excluded by $Z\gamma$ searches at LHC-8TeV.

The other benchmark points are all within the allowed experimental bounds, both from collider and from dark matter experiments,
and can provide viable scenarios to accommodate the di-photon excess as well as to account for the correct relic density of dark matter.
Note that the benchmark points predict a direct detection cross section which is not far from the actual experimental reach, and  will likely be accessible in future experiments.

\begin{center}
\begin{table}[t]
\footnotesize
\begin{tabular}{cccccccccc}
&&  
		  & $\sigma_{\gamma Z}$ 
		  & $\sigma_{\mathrm{MET}+j}  $ 
		  & $ \sigma_{\gamma \gamma}$
		  &$\sigma_{jj}$ 
		  &$ \langle \sigma v  \rangle_{\gamma \gamma} $
		  & $\sigma_{SI}$ \\ \hline
      $(g_{GG} , g_{BB})$   & $M_{\psi}$ [GeV] & $g_{DM}$      &    $< 3.5$ fb  & $< 6 $ fb & $< 2 $ fb& $< 10^3 $ fb & $< 10^{-28} \frac{\mathrm{cm}^3}{\mathrm{s}}$ & $ < 4 \times 10^{-45} \mathrm{cm}^2$ \\ \hline
(0.25,1)& 322 & 2.7
& 0.86 &  3.7 & 1.4  &  1.3 &   $ 3.9 \cdot 10^{-32} $& $6.9 \cdot 10^{-46}$ \\
(0.25,2)& 307 & 2.2
&\alb{3.6} &  3.5 & \alb{6.0} &  1.4 & $5.5 \cdot 10^{-32}$ & $4.6 \cdot 10^{-46}$ \\
(0.14,1)& 323 & 2.7
& 0.3 &  1.2 &  0.48 &  0.14 & $4.1 \cdot 10^{-32} $&  $2.3  \cdot 10^{-46}$  \\
(0.14,2)& 308 & 2.3
& 1.1 & 1.2 & 1.8  & 0.13 & $ 6.2 \cdot 10^{-32} $&  $ 1.6 \cdot 10^{-46}$  \\
\hline

\end{tabular}
\caption{Summary of experimental constraints 
from LHC8 searches and from dark matter experiments
on the four benchmark points described in (\ref{benchmarks}).
All collider cross sections are given in fb and assume $\sqrt{s} = 8 \tev$. For the constraints on $\sigma_{jj}$ we compute the cross section imposing a cut of  $p_T^j > 20 \gev, \eta_j < 2.5$, while for the $\sigma_{\mathrm{MET}+j}$ we impose a cut of $ p_T^j > 500 \gev$. 
}
\label{tab:constraints}
\end{table}
\end{center}

\section{Future signatures}\label{sec4}

In this note we have proposed a simplified dark matter model  with a mediator of mass $\sim 750$ GeV  to account for the di-photon excess recently reported by the ATLAS and CMS
collaborations.
If the resonance is a scalar singlet, the requirement of a moderately large resonance width from the ATLAS collaboration (see (\ref{feature3})) hints to the existence of extra decay channels.. Here we have investigated the possibility that the scalar singlet has an extra decay mode into an
\emph{invisible} particle which can play the role of a dark matter candidate. This simple assumption, together with the requirement of a correct relic abundance,
provides a prediction for the mass of the dark matter, that should be around $\approx300$ GeV for a scalar mediator of $750$ GeV.

The model is indeed very predictive and we can identify the expected signatures in other channels at LHC-13TeV 
and in dark matter experiments.
The most distinctive LHC signature is in jet+MET which has the largest cross section and will be reachable at LHC-13TeV with more luminosity. 
From the model independent analysis of CMS \cite{Khachatryan:2014rra} one can estimate the luminosity needed to exclude our model at 13 TeV by assuming that the efficiencies for the main SM backgrounds are the same as in the 8 TeV run. We focus on the  $\text{MET}>500$ GeV bin, which gave the most stringent constraints at 8 TeV. A back of the envelope estimate indicates that the benchmark point $p_1$ (with large jet+MET cross section)
should be within reach with a few fb$^{-1}$  at 13 TeV.
Benchmark points $p_3$ and $p_4$ (with small jet+MET cross section) would instead need 
few tens of fb$^{-1}$ to be excluded. 
We then argue that essentially all the viable portion of parameter space in figure \ref{xsecplot2}  should be withinin the reach of LHC-13TeV with $\lesssim100$ fb$^{-1}$ of luminosity. A more detailed  analysis, which we leave for a future work,  is necessary in order to extract more precise values for luminosity needed to explore the allowed parameter space in our model. 

The final state $Z\gamma$ is also a promising channel. However, note that by tuning the couplings $g_{BB}$ and $g_{WW}$ one can
generically suppress this branching ratio\footnote{It scales like $\sim (g_{BB}-g_{WW})^2$.}, and thus the signal. Hence the $Z\gamma$ is not a generic prediction of our model, in contrast to jet+MET.

Interestingly, dark matter experiments are also going to probe our model. The future direct detection experiments should reach a sensitivity of approximately $10^{-46}$~cm$^2$ for spin independent cross section assuming a dark matter mass
of around $300$ GeV (see e.g. XENON1T prospects \cite{Cushman:2013zza}), which is in the ballpark of the predictions for our benchmark points 
(see Table \ref{tab:constraints}). In fact, direct detection experiments should be able to probe most of the parameter space of the model which featutres a large width and
is compatible with LHC-8TeV jet+MET searches (i.e. the region illustrated in figure \ref{xsecplot2} with $g_{GG} \lesssim 0.3$),
as the direct detection cross section is set essentially by $g_{GG}$ and $g_{DM}$ (see eq. \eqref{DDsigma}).

Note that both jet+MET and direct detection DM cross sections can be reduced by decreasing the value of the coupling $g_{GG}$. However, in order to maintain
a significant yield in the $\gamma \gamma$ channel, this should be accompanied by an increase of $g_{BB}$, pushing the model into a somehow less appealing region of the parameter space (especially from the point of view of the UV completion). 

We conclude that the simplified dark matter model we presented here provides sharp phenomenological predictions that  can be further scrutinized in both LHC-13TeV and in future searches for galactic dark matter. We leave a more complete exploration of the parameter space of this scenario and the possibility of embedding it into UV complete models beyond the Standard Model for future investigations.

\section*{Acknowledgements}
We thank Brando Bellazzini, Dario Buttazzo, Adam Falkowski, Gilad Perez, Filippo Sala and Michael Spannowsky for interesting discussions. 
We also thank Andreas Goudelis, Thomas Hambye and Paolo Panci for useful comments on the draft.
The work of DR was supported by the ERC Higgs at LHC. M.B. and A.M. are supported in part by the Belgian Federal Science Policy Office through the Interuniversity Attraction
Pole P7/37.  A. M. is supported by the Strategic Research Program High Energy Physics and the Research Council of the Vrije Universiteit Brussel.

\appendix

\section{Comments on UV completion}\label{UVcompl}

In this paper we studied a simplified model of Dirac dark matter with a real scalar mediator described by the effective lagrangian
\begin{align}
\mathcal{L}_{NP}^+&=\frac{1}{2}(\partial S)^2+\frac{m_S^2}{2} S^2+\bar \psi\slashed{\partial}\psi +
(g_{DM} S+M_{\psi})\bar\psi \psi\notag\\
&+\frac{g_{GG}}{\Lambda}S G^{\mu\nu} G_{\mu\nu}+\frac{g_{WW}}{\Lambda}S W^{\mu\nu} W_{\mu\nu}+\frac{g_{BB}}{\Lambda}S B^{\mu\nu} B_{\mu\nu}\ .\label{lagrangianA}
\end{align}
Here below we comment on the assumptions associated with the structure of the lagrangian (\ref{lagrangianA}) 
and the challenges related to their possible UV completions:

\begin{itemize}
\item The dimension five operators of the second line in \eqref{lagrangianA} can be obtained by integrating out  heavy fermionic matter in vector-like representations of the SM gauge group which couples with the singlet $S$ as 
\begin{equation}
\mathcal{L}_{int}=\sum_{i=1}^{N_f}(\lambda_\Psi S+M_\Psi)\bar{\Psi}_i\Psi_i \ .\label{chargedmatter}
\end{equation}
Identifying the cut-off $\Lambda$ with $M_\psi$, we can estimate for a single family: $g_{GG}\approx\frac{\lambda_\Psi\alpha_3Q^2_3(i)d_2(i)}{4\pi}$, $g_{WW}\approx \frac{\lambda_\Psi\alpha_3Q^2_2(i)d_3(i)}{4\pi}$ and $g_{BB}\approx\frac{\lambda_\Psi \alpha_1 Q^2(i)d_3(i)d_2(i)}{4\pi}$ where the $d$'s account for the representation multiplicity in the loop, $Q^2_a$ is the Casimir of the representation of the \emph{a}-th group (which is just $Y^2$ for $U(1)_Y$) and $\alpha_{a}=\tfrac{g_a^2}{4\pi}$ with $a=1,2,3$ are the coupling constants of the SM gauge group.

As we have shown in section \ref{sec3}, having a sizeable yield into di-photons and a sizeable width requires $g_{BB}$ and $g_{DM}$ (and possibly also $g_{GG}$) to be order $\mathcal{O}(1)$. The couplings in front of dimension five operators are loop suppressed in a weakly coupled setup. Having them $\mathcal{O}(1)$ certainly requires some large charge/multiplicity for the vector-like matter which carries SM quantum numbers and/or sizeable couplings of the singlet in \eqref{chargedmatter}. Both these options are likely to induce problems with perturbativity at the UV scale making the UV completions of the effective lagrangian \eqref{lagrangianA} more challenging. 

\item We proceed to comment on other operators allowed by symmetries which we neglected in the effective action \eqref{lagrangianA}. We are neglecting both the cubic and the quartic interaction between the singlet $S$ and the SM Higgs i.e. $\lambda_{3SH}SH^\dagger H$ and $\lambda_{4SH}S^2H^\dagger H$. At the level of dimension five, we are neglecting interactions between the SM Higgs and the Dirac fermion $(g_{HDM} H^\dagger H \bar \psi \psi)$ and also the ones between the singlet and the SM fermions, for instance $g_{SSM} S H Q U$. These operators are not forbidden by any symmetry and should be present in a generic dimension five effective action. 

The presence of singlet-Higgs interactions can modify the decay modes of the singlet which would acquire SM-like couplings through the mixing with Higgs once EWSB is broken. The phenomenology of such a singlet has been widely studied in the literature (see for example \cite{Buttazzo:2015bka}). In our discussion we took as a working assumption (very much in the spirit of \cite{Godbole:2015gma}) that the singlet Higgs interactions are zero at the UV thresholds $\Lambda$. In such a hypothesis we can also approximate the singlet potential with the mass term only, since a singlet vacuum expectation value would not have any consequence on the phenomenology besides modifying the DM mass. Once we set the tree level couplings to zero, the UV threshold corrections coming from loops of vector-like charged matter \eqref{chargedmatter} will generate these couplings only at 2 loops. Therefore the Higgs-singlet couplings would be loop suppressed with respect to the couplings to gauge bosons once the tree level boundary condition is realized.
\end{itemize}
An attractive possibility to get a structure of the couplings at the UV boundary condition fulfilling both the challenges described above can be found in the context of supersymmetry (SUSY). 
In a generic low energy SUSY-breaking model, there is a light pseudo-modulus (i.e. the sgoldstino) which sits on the same SUSY multiplet of the spin 1/2 goldstino associated to spontaneous SUSY-breaking \cite{Komargodski:2009rz}
(see \cite{Dudas:2012fa} for a more detailed discussion of the sgolstino phenomenology at colliders). The structure of the couplings of our effective lagrangian \eqref{lagrangianA}
would describe the interactions between the CP-even component of the sgoldstino and the SM, as well as the invisible decay to the neutral goldstino. In the limit in which gaugino masses are heavy, the dimension five couplings of the sgoldstino to gauge bosons dominate over the other couplings,
since they scale like $\sim M_i/\sqrt{f}$, where $M_i$ are the gaugino masses and $f$ is the SUSY-breaking scale. Couplings of $\mathcal{O}(1)$ can be obtained in extra dimensional scenarios where the gaugino masses are generated at tree level \cite{Chacko:1999mi,Csaki:2001em}. 
In this case, however, the supersymmetry breaking scale will be tied to the gaugino mass and will be bounded from below from constraints on the gluino mass and from above by the requirement of a sizable cross section, imposing some further constraints on the parameters of the effective lagrangian.
For instance, the goldstino would not be a suitable dark matter candidate since for low $\sqrt{f}$ it would result in very light dark matter.

A possibility to include a dark matter candidate in such context is to consider
models of pseudo-moduli dark matter constructed in \cite{Shih:2009he,KerenZur:2009cv,Amariti:2009tu}. In these constructions one or more light chiral superfields generically arise in 
O'Raifeartaigh
models which break SUSY spontaneously. Their scalars components are pseudo-moduli associated to approximately flat directions in the potential while their fermionic component can be a viable dark matter candidate. The lagrangian \eqref{lagrangianA} describes the CP-even component of a complex pseudo-modulus,
coupled to a singlet Dirac-like Dark whose mass is not tight to the SUSY-breaking scale. 
The latter can easily arise in the context of supersymmetry as a pair of Weyl fermions with a small supersymmetric mass which remains light once their scalar partners have acquired a large SUSY-breaking mass.

Note that in SUSY inspired scenarios, we expect  the CP even and CP odd part of the light complex 
pseudo-modulus to have the same mass.
Hence, as a further remark,
we would like to briefly comment on the possibility that the scalar resonance is not CP even but CP odd.
Considering the analogous of the lagrangian (\ref{lagrangianA}) but in the case of a CP odd scalar, the decay rates to SM gauge boson would be equivalent, while the invisible decay would present a 
phase space suppression different than eq. (\ref{invisibleCPeven}).
The analysis of the LHC di-photon excess could proceed
very similar to our study, however the dark matter features are significantly different. 
In the case of a pseudo-scalar, the annihilation 
cross section relevant for indirect detection is not velocity suppressed and gives a large contribution to $\langle \sigma v \rangle_{\gamma \gamma}$.
As a consequence, a dark matter of mass around $250-300$ GeV 
with $O(1)$ coupling to the scalar mediator would lead to a very large yield to indirect detection experiments, 
typically of the order of $\langle \sigma v \rangle_{\gamma \gamma} \sim 10^{-26}\, \frac{\text{cm}^3}{\text{s}}$, 
which would be in tension with the FERMI constraint \cite{Ackermann:2015lka}.
Note that such a large dark matter mass is necessary to 
obtain the correct relic abundance through the resonant enhancement of the annihilation via the $750$ GeV mediator. Hence in the case of
a pseudo-scalar mediator the relic abundance requirement and the indirect detection bounds would generically be in tension. We leave to future studies a detailed investigation of this case.

Finally, let us mention that a similar effective theory to \eqref{lagrangianA} could also arise in the context of Randall-Sundrum scenarios where the light singlet is a dilation/radion of a hidden sector where conformal symmetry is spontaneously broken. However,  the couplings to quarks in these scenarios are typically sizeable and they strongly modify the phenomenology of the singlet (see \cite{Blum:2014jca} for a detailed study of the implications of this setup for dark matter).

\FloatBarrier

\bibliography{biblio}

\end{document}